\documentclass[%
 preprint, 
 amsmath,amssymb,
 aps, physrev,
]{revtex4-2}

\usepackage{graphicx}% Include figure files
\usepackage{dcolumn}% Align table columns on decimal point
\usepackage{bm}% bold math
\usepackage{dutchcal}

\begin{document}

\preprint{}

\title{\textbf{The Role of Drop Shape in Impact Force} 
}% 

\author{Yang Zeng}
\author{Zhen Chen}
\author{Lei Xu}%
 \email{xuleixu@cuhk.edu.hk}
\affiliation{%
The Department of Physics, The Chinese University of Hong Kong, Shatin, NT, Hong Kong.
}%

\date{\today}% It is always \today, today,

\begin{abstract}
Drop impacts are ubiquitous in natural and industrial processes, yet the influence of drop shape on impact force remains a fundamental open question. Combining experiments with theoretical analysis, we show that drop shape plays a critical role, with impact force varying by more than an order of magnitude solely due to changes in shape. By uncovering self-similarity in time and cross-shape similarity across diverse drop profiles, we develop a universal cylinder model that accurately predicts both the magnitude and timing of the impact force. This study establishes a comprehensive framework for understanding impact forces across a wide range of drop shapes. Given the prevalence of drop impacts with varying shapes in real-world scenarios, our findings hold fundamental significance and have broad potential applications across industries such as soil erosion prevention, jet cutting, spray coating, and design of windshields and wind turbines.
\end{abstract}

\maketitle

%\tableofcontents
\section{\label{sec:level1}Introduction}
The impact of liquid drops on solid surfaces is a common phenomenon observed in both natural and industrial processes \cite{yarin2006drop,josserand2016drop}, including soil erosion \cite{cheng2022drop}, surface coating \cite{pan2018coatings}, inkjet printing \cite{lohse2022fundamental}, spray cooling \cite{jiang2022inhibiting}, and respiratory transmission \cite{bourouiba2021fluid}. Recent advancements in high-speed photography have facilitated detailed investigations into key outcomes of drop impact, such as air entrapment \cite{liu2013compressible, lee2012does}, spreading \cite{roisman2009inertia, lagubeau2012spreading}, splashing \cite{xu2005drop, liu2015kelvin, xu2010instability, xu2007liquid}, and rebound \cite{richard2002contact}. While extensive research has been conducted on drop impact \cite{yarin2006drop, josserand2016drop, cheng2022drop, pan2018coatings, lohse2022fundamental, jiang2022inhibiting, bourouiba2021fluid, liu2013compressible, lee2012does, roisman2009inertia, lagubeau2012spreading, xu2005drop,liu2015kelvin, richard2002contact}, these studies predominantly focus on spherical drops, with relatively little attention given to non-spherical drops.

In real-world scenarios, however, drops often exhibit various non-spherical shapes, such as raindrops deformed by air drag \cite{kostinski2009raindrops}, charged drops in electric fields \cite{ristenpart2009non, beroz2019stability}, ferrofluid drops in magnetic fields \cite{fan2020reconfigurable}, and oscillating drops \cite{beard1989natural}. Therefore, studying the impact of non-spherical drops not only enhances our fundamental understanding of drop impact but also holds significant potential for practical applications, given its relevance to numerous real-world and industrial processes \cite{liu2021role}.

Furthermore, most experimental studies have primarily focused on the kinematic aspects of drop impact \cite{yarin2006drop, josserand2016drop}, as these features are easily captured and analyzed using high-speed photography. However, dynamic features, such as impact force \cite{gordillo2018dynamics} and internal force distribution \cite{sun2022stress}, have received significantly less attention until the recent development of high-precision piezoelectric force sensors \cite{li2014impact}. These dynamic features are crucial in processes such as soil erosion, jet cutting, and the design of durable windshields and wind turbines \cite{zhao2015granular, field1999elsi}. Additionally, current investigations have again concentrated on spherical drops \cite{cheng2022drop, gordillo2018dynamics, sanjay2025unifying, zhang2022impact, mitchell2019transient}, leaving a significant knowledge gap regarding the impact force of non-spherical drops. Consequently, a universal understanding of impact forces across various drop shapes remains elusive.

To develop a universal understanding, we experimentally produce various non-spherical drops and examine how drop shape influences impact force and internal force distribution. Non-spherical drops are described by the superellipse equation: $|x/a|^n+|y/b|^n=1$, where $a$ and $b$ represent the global feature of semi-horizontal and semi-vertical axes, and $n$ is the power index indicating the local curved profile. Our findings reveal that both the global aspect ratio ($\alpha=a/b$) and the local feature of $n$ fundamentally affect the impact force. For drops with the same impact momentum, oblate drops with large $\alpha$ and $n$ can exhibit a maximum impact force $10$ times higher than prolate ones! Furthermore, we identify a universal behavior in the impact force, characterized by self-similarity across different moments and cross-shape similarity across various drop shapes. Building on this, we develop a general cylinder model to predict the time and magnitude of maximum impact force, which agrees well with experimental results across different drop shapes. Our study offers a universal and fundamental understanding of impact forces across various drop shapes. This understanding is not only crucial for advancing fundamental research but also holds significant practical value in a wide range of drop-impact-related industrial processes.

\section{\label{sec:level1}Experimental Results}
The experimental setup is illustrated in Fig. 1a. A free-falling ferrofluid drop first passes through a magnetic coil, which generates a strong magnetic field, stretching the drop into a long spindle-like shape. The magnetic field is then rapidly turned off before the drop impacts on the surface, allowing the drop to oscillate under surface tension. The timing of the magnetic field’s deactivation is precisely controlled using a laser trigger and an off-delay timer. By carefully adjusting the turn-off time of the magnetic field, we can produce various drop shapes at the moment of impact \cite{liu2021role}. Fig. 1b presents examples of drop cross-sectional shapes, which, in 3D, are axisymmetric around the vertical central axis. These cross-sectional shapes can be described by the superellipse equation: $|x/a|^n+|y/b|^n=1$, where $a$ and $b$ are the semi-horizontal and semi-vertical axes, and $n$ is the superellipse power index ($n=2$ for ellipses and spheres). The aspect ratio ($\alpha=a/b$) ranges from $0.5$ to $2.2$ and $n$ ranges from $1.8$ to $2.8$ in our experiments.

Detailed experimental parameters are listed below. The ferrofluid has the density $\rho=1.2 \times 10^3 \ \rm kg/m^3$, the dynamic viscosity $\mu=8 \ \rm cP$, and the surface tension $\gamma=19 \ \rm mN/m$ (at $20 \ \rm ^\circ C$ and in the absence of magnetic field). The impact velocity is $V_0=2.6 \pm 0.1 \ \rm m/s$, the initial spherical drop diameter is $D=2.9 \pm 0.1 \ \rm mm$, the Reynolds number is $Re=\rho DV_0/\mu=1110 \pm 50$, and the Weber number is $We=\rho V_0^2D/\gamma=1260 \pm 80$.

Fig. 1c presents selected snapshots of superelliptical drops impacting a solid surface, and Fig.1d shows the corresponding impact force measurements. The impact force and time can be expressed in dimensionless forms: $F^*=F/(\rho(2a)^2 V_0^2)$, and $t^*=t/(2b/V_0)$. Clearly, the force curves exhibit significant differences in both magnitude and profile (see Supplemental Movie 1-3 \cite{SI_DropImpact}). The dimensionless maximum force $F_{\rm max}^*$ systematically increases as the drop shape transitions from prolate to oblate. Moreover, the maximum force of a prolate drop shows a plateau rather than a well-defined peak, and the exact location of $F_{\rm max}^*$ on this plateau is taken as the beginning of the plateau (see SM Sec. 3 \cite{SI_DropImpact}).

\begin{figure}
    \centering
    \includegraphics[width=0.75\linewidth]{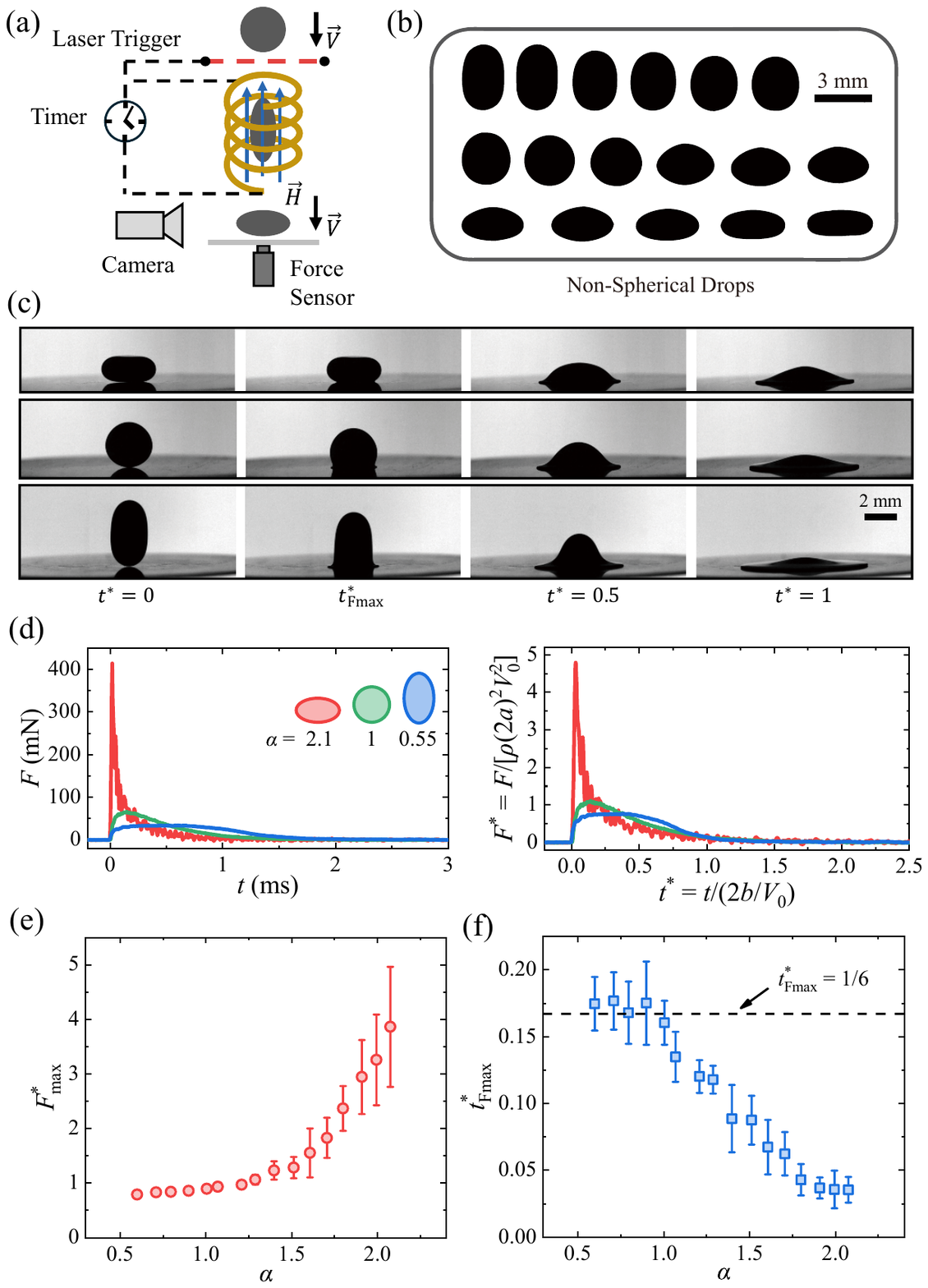}
    \caption{Experimental setup and impact force measurements. (a) The experimental setup for producing non-spherical drops. (b) Cross-sections of drops with various shapes generated by our method. (c) Snapshots of drop impacting on a solid surface by oblate (top), spherical (middle), and prolate (bottom) drops. (d) The impact force (left panel) and its dimensionless form (right panel) for oblate, spherical, and prolate drops measured by force sensor. (e) The dimensionless maximum force $F_{\rm max}^*$ as the function of the aspect ratio $\alpha$. (f) The dimensionless time to reach the maximum force $t_{\rm Fmax}^*$ as the function of $\alpha$.}
    \label{fig1}
\end{figure}

A systematic measurement of $F_{\rm max}^*$ as a function of the aspect ratio $\alpha$ is shown in Fig. 1e, clearly demonstrating that a more oblate shape with a larger $\alpha$ results in a greater $F_{\rm max}^*$. More surprisingly, in the dimensional form $F_{\rm max}$ varies from $34 \ \rm mN$ (at $\alpha=0.55$) to $420 \ \rm mN$ (at $\alpha=2.1$) (see the left panel in Fig. 1d), spanning an entire order of magnitude solely due to the shape change! This unambiguously highlights the significant role of drop shape in the impact force.

Besides the maximum force ($F_{\rm max}^*$), we further investigate the time to reach this maximum, $t_{\rm Fmax}^*$, at different aspect ratios. As shown in Fig. 1f, $t_{\rm Fmax}^*$ remains relatively constant around $1/6$ when $\alpha\le1$, but decreases systematically when $\alpha > 1$. Notably, $t^*= 1/6$ corresponds to the specific moment when the contact radius equals the horizontal drop radius ($r_V=a$) (see derivations in SM Sec. 4 \cite{SI_DropImpact}), as illustrated by the snapshots in Fig. 1c column 2, row 3. The contact radius is determined by the region directly beneath the main body of the drop, excluding the tip portion due to its negligible contribution to the impact force (see Fig. 1c). This result ($t^*_{\mathrm{Fmax}}=1/6$) aligns with previous studies on spherical drops \cite{mitchell2019transient}.

However, this previously observed behavior does not apply to oblate drops. For these drops,  $t_{\rm Fmax}^*$ can be as small as 0.03 (i.e., 5 times shorter), as shown in Fig. 1f. The fundamental difference will be illustrated by our cylinder model later: for spherical and prolate drops, $F_{\rm max}^*$ and $t_{\rm Fmax}^*$ are determined by the side boundary of the drop, whereas for oblate drops the top boundary becomes the determining factor.

\section{\label{sec:level1}Theoretical Analysis}

\textbf{\textit{1. Self-Similarity of the Impact Force at Different Times.}} To analyze the impact force, we examine the force density (i.e., force-per-unit-volume) distribution within the drop through simulations (see SM Sec. 2 \cite{SI_DropImpact}). Specifically, we focus on the vertical force produced by the pressure gradient along the $z$-axis, $f_p (r,z)=-\partial p(r,z)/\partial z$, whose integration over the drop volume gives the total impact force, $F_{\rm tot}=\int f_p \mathrm{d}\Omega$. 

We begin with the simplest case: a spherical drop. As shown in Fig. 2a (top row), at early time ($t^*=1/48$) a region of high $f_p$ is concentrated near the contact area, forming an approximately cylindrical shape. As time progresses ($t^*\ge 1/24$), a region of low $f_p$ emerges from the bottom of the drop, as a result of the loss of impact momentum near the bottom area. Additionally, the magnitude of $f_p$ decreases over time, reflecting the overall reduction of impact momentum.

However, once the length and force are rescaled using appropriate scaling factors, the distribution of $f_p$ exhibits a self-similar behavior across different times \cite{philippi2016drop}. Since the contact radius of a spherical drop spreads as $\tilde{r}_V\propto\sqrt{\tilde{t}}$ at early stage \cite{liu2021role}, the length scale can be rescaled by $\sqrt{\tilde{t}}$, leading to the rescaled coordinates: $\xi=\tilde{r}/\sqrt{\tilde{t}}$ and $\eta=\tilde{z}/\sqrt{\tilde{t}}$, where $\tilde{r}=r/R$, $\tilde{z}=z/R$, $\tilde{t}=t/(R/V_0)$ and $R$ is the radius of sphere. Note that quantities with $\tilde{}$ are dimensionless quantities specifically introduced for the rescaling analysis, which may differ from dimensionless quantities with $^*$ that are previously defined.

Correspondingly, the force can also be rescaled with time as: $\mathcal{f}_\mathcal{P} (\xi,\eta)=\tilde{t}\tilde{f}_p (r,z)$ (see SM Sec. 5 \cite{SI_DropImpact}), where $\tilde{f}_p=f_p/(\rho V_0^2/R)$ with $f_p$ the actual force density. After applying these rescaling, the rescaled force density $\mathcal{f}_\mathcal{P} (\xi,\eta)$ exhibits nearly identical distributions under rescaled length scales $\xi$ and $\eta$ across different moments, as shown in Fig. 2a bottom row. This self-similar behavior remains valid at $t^*=1/6$, until the top surface of the drop approaches and destroys it at $t^*=1/3$ (see Fig. 2a bottom row).

\textbf{\textit{2. Cross-Shape Similarity of the Impact Force for Different Drop Shapes.}} The similarity analysis conducted for spherical drops can be extended to elliptical drops described by the equation: $|x/a|^2+|y/b|^2=1$. For these elliptical drops, the growth of the contact radius follows the same scaling relationship \cite{liu2021role}, $\tilde{r}_V\propto\sqrt{\tilde{t}}$, where $\tilde{r}=r/R_c$, $\tilde{t}=t/(R_c/V_0)$, with $R_c=a^2/b$ serving as the characteristic length scale (it is the local radius of curvature at drop bottom). By applying the same procedure as used for a spherical drop, but replacing $R$ with $R_c$, the impact force distribution $f_p (r,z)$ can again be rescaled to $\mathcal{f}_\mathcal{P} (\xi,\eta)$, where $\xi=\tilde{r}\sqrt{\tilde{t}}$ and $\eta=\tilde{z}/\sqrt{\tilde{t}} $ (see details in SM Sec. 5 \cite{SI_DropImpact}).

The original $f_p (r,z)$ and rescaled $\mathcal{f}_\mathcal{P} (\xi,\eta)$ are again illustrated in top and bottom rows of Fig. 2b, for elliptical drops spanning a broad range of aspect ratios, $0.5\le a/b\le 2.2$. The time is chosen at the early impact stage ($t^*<1/48$). Notably, despite the significant variations in $f_p (r,z)$ shown in the top row, the bottom row shows nearly identical distributions of $\mathcal{f}_\mathcal{P} (\xi,\eta)$. This result confirms excellent cross-shape similarity, from prolate to oblate elliptical drops.

Moreover, we can extend the similarity analysis even further to superelliptical drops, which are described by the equation: $|x/a|^n+|y/b|^n=1$. With the additional parameter $n$, the superellipse generalizes the ellipse shape (where $n=2$) and can describe various drop shapes generated in our experiments. For these superelliptical drops, the early-stage spreading of the contact radius can be described as\cite{liu2021role}:
\begin{equation}
\tilde{r}_V=(n(n+2)/2)^{1/n} (\tilde{t})^{1/n}
\end{equation}
where $\tilde{r}_V=r_V/R_{c,n}$, $\tilde{t}=t/(R_{c,n}/V_0)$, $R_{c,n}=R_c(a/b)^\beta$, $R_c=a^2/b$, and $\beta=(2-n)/(n-1)$ (see derivation details in SM Sec. 5 \cite{SI_DropImpact}). Note that the new characteristic length scale of superellipse is $R_{c,n}$, which is not the local radius of curvature at the bottom of superellipse \cite{stone2005lubrication}.

\begin{figure}
    \centering
    \includegraphics[width=0.75\linewidth]{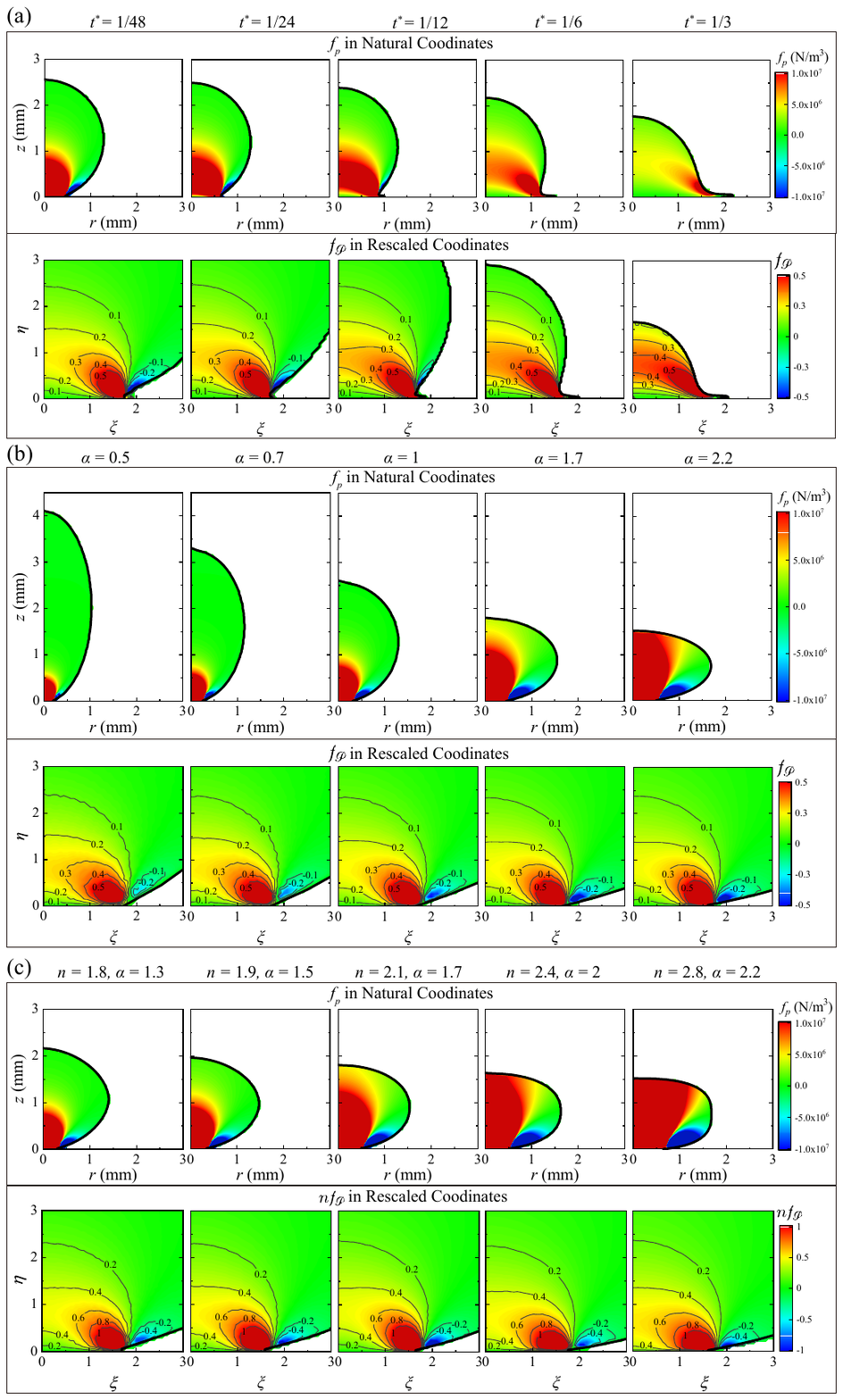}
    \caption{The self-similarity and cross-shape similarity of the impact force distribution. (a) The evolution of the force density distribution calculated from pressure gradient along $z$-axis, denoted as $f_p (r,z)$ in the natural coordinates (top), whose rescaled representation $\mathcal{f}_\mathcal{P} (\xi,\eta)$ exhibits self-similar behavior (bottom). (b) The force density distribution $f_p (r,z)$ of different elliptical drops at early stage (top), whose rescaled representation $\mathcal{f}_\mathcal{P} (\xi,\eta)$ exhibits cross-shape similarity across different $\alpha$ (bottom). (c) The force distribution $f_p (r,z)$ of different superelliptical drops at early impact stage (top), whose rescaled representation $n\mathcal{f}_\mathcal{P} (\xi,\eta)$ exhibits cross-shape similarity across different $\alpha$ and $n$. (bottom).}
    \label{fig2}
\end{figure}

Unlike elliptical drops, where the spreading radius scales as $\tilde{r}_V\propto (\tilde{t})^{1/2}$, for superelliptical drops the spreading radius scales as $\tilde{r}_V\propto(\tilde{t})^{1/n}$. This introduces fundamental changes to the similarity representation. The rescaled coordinates must now be reformulated as: $\xi=\tilde{r}/(\tilde{t})^{1/n}$, $\eta=\tilde{z}/(\tilde{t})^{1/n}$, and the rescaled force $\mathcal{f}_\mathcal{P}(\xi,\eta)$ scales similarly as the elliptical case:
\begin{equation}
\mathcal{f}_\mathcal{P}(\xi,\eta)=\tilde{t}\tilde{f}_p (\tilde{r},\tilde{z})
\end{equation}
where $\tilde{r}=r/R_{c,n}$, $\tilde{z}=z/R_{c,n}$ and $\mathcal{f}_p=f_p/(\rho V_0^2/R_{c,n})$. 

However, we find that $\mathcal{f}_\mathcal{P} (\xi,\eta)$ is no longer the invariant quantity and the product $n\mathcal{f}_\mathcal{P}(\xi,\eta)$ becomes the new invariant quantity (see derivation details in SM Sec. 5 \cite{SI_DropImpact}), i.e.:
\begin{equation}
n\mathcal{f}_\mathcal{P}(\xi,\eta)=C(\xi,\eta)
\end{equation}
where $C(\xi,\eta)$ is a fixed distribution function that remains unchanged regardless of shape variations. Since Eq.(3) holds across different powers of $n$ and aspect ratios $\alpha$ (including spherical and elliptical cases), it represents a general similarity relationship across different drop shapes. By contrast, the previously observed similarity of $\mathcal{f}_\mathcal{P} (\xi,\eta)$ for spherical and elliptical cases is a special instance due to a fixed $n$ ($n=2$). In more general cases with varying $n$, the invariant quantity should be replaced by $n\mathcal{f}_\mathcal{P}(\xi,\eta)$.

To verify Eq. (3), we illustrate original $f_p (r,z)$ and rescaled $n\mathcal{f}_\mathcal{P} (\xi,\eta)$ for various $n$ and $\alpha$ in Fig. 2c (top and bottom rows). The bottom row shows excellent cross-shape similarity of $n\mathcal{f}_\mathcal{P} (\xi,\eta)$ for a wide range of drop shapes with different $n$ and $\alpha$, despite the significant differences in $f_p (r,z)$ appeared in the original coordinates (top row). 

Combining the self-similarity results in Fig. 2a with the cross-shape similarity results in Fig. 2b and 2c, our study uncovers both self-similarity across different moments and cross-shape similarity across various drop shapes. This represents the most general similarity scenario in drop impact research to date.

\textbf{\textit{3. A Simple Yet Universal Cylinder Model.}} Based on the self-similarity and cross-shape similarity properties, we analyze the detailed distribution of $\mathcal{f}_\mathcal{P} (\xi,\eta)$ and propose a simple but universal cylinder model. Since different shapes exhibit similarity, we can simplify the analysis by focusing on the spherical shape. To do so, we take a snapshot of the spherical drop from Fig. 2a and analyze it in Fig. 3a.

\begin{figure}
    \centering
    \includegraphics
    [trim=10 30 10 30,clip, width=0.70\linewidth]{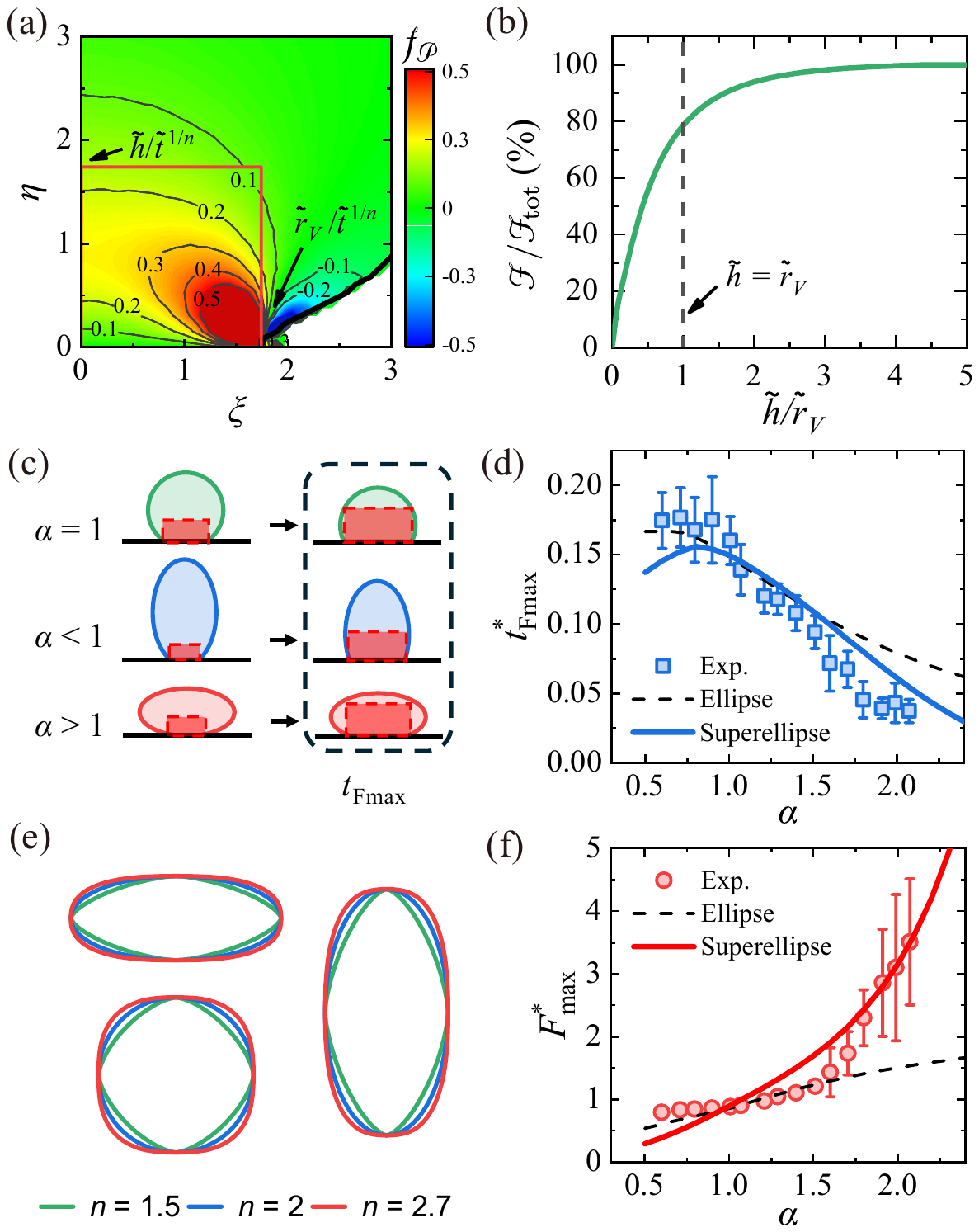}
    \caption{The cylinder model. (a) The simulation of rescaled force $\mathcal{f}_\mathcal{P}$ for a spherical drop ($n=2$). The high-$\mathcal{f}_\mathcal{P}$ cylindrical region is marked by the red square. (b) The impact force integrated within a cylinder ($\mathcal{F}$) with varying heights ($\tilde{h}$) divided by the total impact force $\mathcal{F}_{\rm tot}$. (c) Schematics illustrating the growth of the cylinder to reach maximum volume for drops with different aspect ratios. (d) Comparison between the cylinder model prediction and experimental data on $t_{\rm Fmax}^*$. The dashed curve represents the elliptical shapes at fixed $n=2$ and the solid curve represents the superelliptical shapes with different $\alpha$ and $n$. (e) The schematics of superellipse demonstrates a flatter drop profile for a larger n across different aspect ratio $\alpha$. (f) The prediction on $F_{\rm max}^*$ by the cylinder model. The dashed curve represents the elliptical shapes with fixed $n=2$ and the solid curve represents the superelliptical shapes with different $\alpha$ and $n$.}
    \label{fig3}
\end{figure}

It is evident that the high-$\mathcal{f}_\mathcal{P}$ region is primarily concentrated near the bottom, forming an approximately cylindrical shape (see the 0.2 contour of Fig. 3a, for example). This region makes the most significant contribution to the overall impact force.

Therefore, we can approximate the total impact force by integrating $\mathcal{f}_\mathcal{P}$ within this cylindrical region. The radius of this cylinder is simply the contact radius of the drop, $\tilde{r}_V$. For the cylinder’s height, the simplest assumption is to set it equal to the radius, $\tilde{h}=\tilde{r}_V$. 

To test this assumption, we integrate $\mathcal{f}_\mathcal{P}$ for cylinders of varying heights $\tilde{h}$, as shown in Fig. 3b. Compared with the total impact force $\mathcal{F}_{\rm tot}$, the force within the cylinder increases sharply with height $\tilde{h}$ and reaches approximately 80\% of $\mathcal{F}_{\rm tot}$ when $\tilde{h}=\tilde{r}_V$. Beyond this point, the increase is much slower, eventually saturating at 100\% of $\mathcal{F}_{\rm tot}$. Thus, 80\% of $\mathcal{F}_{\rm tot}$ demonstrates that the simple assumption of $\tilde{h}=\tilde{r}_V$ effectively defines a high-$\mathcal{f}_\mathcal{P}$ cylindrical region, which captures the majority of the impact force. 

This assumption is not only valid for spherical drops but also holds for elliptical and superelliptical drops, due to the cross-shape similarity. Interestingly, assumptions similar to $\tilde{h}=\tilde{r}_V$ have been made in previous research on water wave impacts, where the impact depth is approximated by the half of the impact width \cite{bagnold1939interim, cooker1995pressure}.

Thus, we can construct a simple cylinder model to explain the impact force by considering a cylinder with radius $\tilde{r}_V$ (contact radius) and height $\tilde{h}=\tilde{r}_V$. This model provides a simple but effective representation of the primary region contributing to the impact force.

Based on this model, we proceed to calculate the maximum force, $F_{\rm max}^*$, and the time to reach the maximum force, $t_{\rm Fmax}^*$, and compare them with actual measurements. Since the impact force distribution is self-similar and primarily determined by the force within the cylinder, the larger the cylinder volume, the greater the impact force. Therefore, the initial expansion of the contact radius and, consequently, the cylinder volume leads to a dramatic increase in impact force. The force reaches its maximum when the cylinder attains its maximum volume within the drop, which is constrained by the drop boundary.

Fig. 3c illustrates the evolution of the cylinder for drops with different aspect ratios. For prolate and spherical drops ($\alpha \le 1$), the maximum volume is reached when the cylinder expands to approach and touch the side boundary of the drop (i.e., $r_V\approx a$, see the spherical and prolate cases in the dashed box of Fig. 3c). This configuration corresponds to the drop’s side boundary being approximately perpendicular to the substrate, typically occurring around $t^*=1/6$. In contrast, for oblate drops ($\alpha>1$), the cylinder may approach and touch the top boundary of the drop first, before expanding to the side boundary (see the oblate plot in the dashed box). As a result, the maximum force is reached much earlier than $t^*=1/6$. These analyses consistently explain the high-speed images captured at $t_{\rm Fmax}^*$ for oblate, spherical and prolate drops in Fig.1c.

With this simple picture, we can obtain the time $t_{\rm Fmax}^*$ directly from the drop profile by determining the moment when the cylinder reaches its side or top boundary (see SM Sec. 6 \cite{SI_DropImpact}). We then compare our theoretically calculated $t_{\rm Fmax}^*$ with the measurements from the force sensor, as shown in Fig. 3d. Without any fitting parameter, our calculation based on elliptical profiles for $n=2$ (dashed black curve) shows reasonable agreement with the experimental data.

However, there is a systematic deviation, particularly for large $\alpha$, as our drops may exhibit powers different from $n=2$ — i.e., a superellipse rather than a standard ellipse profile. By incorporating the actual $n$ values measured from high-speed images, we obtain the solid curve, which aligns more closely with the experimental data. This indicates that both the aspect ratio ($\alpha$) and the power index ($n$) of drop shape play critical roles in determining $t_{\rm Fmax}^*$.

The influence of $n$ is more clearly illustrated in Fig. 3e: across different $\alpha$ values, which indicates the global shape, a larger $n$ results in a locally flatter drop profile, leading to a faster expansion (i.e., $\tilde{r}_V\propto(\tilde{t})^{1/n}$  for small $\tilde{t}$) and an earlier $t_{\rm Fmax}^*$. Additionally, $n$ also affects the magnitude of $F_{\rm max}^*$ significantly: a flatter drop profile with a larger $n$ provides more space to accommodate a larger cylinder, resulting in a larger impact force $F_{\rm max}^*$, as verified next.

Next, we calculate the maximum force, $F_{\rm max}^*$, using the cylinder model and compare it with experiment. Under rescaled coordinates, the force within the cylinder can be expressed as: $\tilde{F}=\int\tilde{f}_p \rm{d}\tilde{\Omega}$, where the integration is over the cylinder volume, $\tilde{\Omega}_{\mathrm{c}}=\pi \tilde{r}_V^2h=\pi \tilde{r}_V^3$. More specifically: $\tilde{F}=\int \tilde{f}_p\mathrm{d}\tilde{\Omega}=\int(\mathcal{f}_\mathcal{P}/\tilde{t})\mathrm{d}\tilde{\Omega}=\int( n \mathcal{f}_\mathcal{P})\mathrm{d}\tilde{\Omega}/(n\tilde{t})=\overline{n\mathcal{f}_\mathcal{P}} \tilde{\Omega}_{\mathrm{c}}/(n\tilde{t})$, where $\overline{n\mathcal{f}_\mathcal{P}}=\int n\mathcal{f}_\mathcal{P}\mathrm{d}\tilde{\Omega}/\tilde{\Omega}_{\mathrm{c}}$ is the average value of $n\mathcal{f}_\mathcal{P}$ within the cylinder. 

From the cross-shape similarity described in Eq.(3), we know that $n\mathcal{f}_\mathcal{P}$ is invariant and thus ($\overline{n\mathcal{f}_\mathcal{P}}$) remains constant across different drop shapes. Therefore, the maximum force is simply proportional to the maximum cylinder volume $\tilde{\Omega}_{\rm c,max}$ that can be accommodated by the drop:
\begin{equation}
\tilde{F}_{\mathrm{max}}=\frac{1}{n}\frac{\tilde{\Omega}_{\mathrm{c, max}}}{\tilde{t}_{\mathrm{Fmax}}}
\end{equation}

By first obtaining $\tilde{F}_{\rm max}$ in rescaled coordinates and then converting to natural coordinates, we can calculate the maximum impact force, $F_{\rm max}^*$, from the cylinder model. A single fitting pre-factor is introduced to account for both ($\overline{n\mathcal{f}_\mathcal{P}}$) and the small contribution from the volume beyond the cylinder (see SM Sec. 6 \cite{SI_DropImpact}).

The comparison between the cylinder model calculation (solid curve) and the measurements from the force sensor is shown in Fig. 3f. With just one fitting parameter, the model demonstrates strong agreement with experimental data across a variety of drop shapes, validating the robustness and general applicability of the cylinder model.

For comparison, we also plot the elliptical situation as a dashed curve, which only considers the case of fixed $n=2$. A systematic deviation is evident, highlighting that both $n$ and $\alpha$ are crucial, and their combination provides the most comprehensive and robust model.

There are still minor deviations between the model (solid curve) and the measurements, primarily due to the breakdown of the power-law expansion assumption, which is strictly valid only as time approaches zero. Nonetheless, the model's simplicity, coupled with its ability to provide remarkably accurate predictions, establishes it as an excellent framework for understanding the essential physics of impact force across diverse drop shapes.

\section{\label{sec4}Conclusion}
In this work, we study the impact force of drops with various shapes described by superellipse equation. Our findings reveal that both the aspect ratio ($\alpha=a/b$) and the power index $n$ fundamentally influence the impact force. For drops with the same impact momentum, oblate drops with large $\alpha$ and $n$ can exhibit a maximum impact force up to 10 times higher than prolate ones.

Furthermore, we identify a universal behavior in the force density distribution, characterized by self-similarity across different times and cross-shape similarity across various drop shapes. Building on these principles, we develop a general cylinder model to predict $F_{\mathrm{max}}^*$ and $t_{\mathrm{Fmax}}^*$, which shows strong agreement with experimental results across a wide range of drop shapes.

Our study provides a universal and fundamental understanding of impact forces across various drop shapes, offering valuable insights into the fundamental physics of drop impacts and shedding new light on applications in a variety of relevant industries.

\begin{acknowledgments}
L.X. acknowledges the financial support from Hong Kong RGC GRF 14305025 and CUHK direct grant 4443366 and 4442847.
\end{acknowledgments}

\bibliography{Refences}% Produces the bibliography via BibTeX.

\end{document}